\documentclass[12pt,preprint]{aastex}
\usepackage{amssymb}
\usepackage{epsfig}
\usepackage{graphicx}
\usepackage{subfigure}
\usepackage{multirow}
\usepackage{tabularx}
\usepackage{rotating}

\newcommand       \eV           {\,{\rm eV}}
\newcommand       \keV          {\,{\rm keV}}
\newcommand       \g            {\,{\rm g}}
\newcommand       \cm           {\,{\rm cm}}
\newcommand       \s            {\,{\rm s}}
\newcommand       \photons      {\,{\rm photons}}
\newcommand       \EWR          {\gamma_{\rm Cr/Fe}}
\newcommand       \EWFe         {\rm EW(Fe)}
\newcommand       \EWCr         {\rm EW(Cr)}
\newcommand       \MCr          {\rm M(Cr)}
\newcommand       \MFe          {\rm M(Fe)}
\newcommand       \dof          {\rm d.o.f}
\newcommand       \simali      {\,\sim}

\begin{document}
\title{
       Cr-K Emission Line as a Constraint on
       the Progenitor Properties of Supernova Remnants
       }
\author{X.J. Yang\altaffilmark{1,2},
        H. Tsunemi\altaffilmark{3},
        F.J. Lu\altaffilmark{4},
        Aigen Li\altaffilmark{2},
        F.Y. Xiang\altaffilmark{1,2},
        H.P. Xiao\altaffilmark{1}, and
        J.X. Zhong\altaffilmark{1}
        }
\altaffiltext{1}{Department of Physics,
                 Xiangtan University,
                 Xiangtan 411105, China;
                 {\sf xjyang@xtu.edu.cn}}
\altaffiltext{2}{Department of Physics and Astronomy,
                 University of Missouri,
                 Columbia, MO 65211, USA
                 }
\altaffiltext{3}{Department of Earth and Space Science,
                 Osaka University,
                 Osaka 560-0043, Japan
                 }
\altaffiltext{4}{Key Laboratory for Particle Astrophysics,
                 Institute of High Energy Physics,
                 Chinese Academy of Sciences,
                 Beijing 100049, China
                 }

\begin{abstract}
We perform a survey of the Cr, Mn and Fe-K emission lines
in young supernova remnants (SNRs)
with the Japanese X-ray astronomy satellite {\sl Suzaku}.
The Cr and/or Mn emission lines are detected in 3C\,397 and 0519-69.0
for the first time. We also confirm the detection of these lines
in Kepler, W49B, N103B and Cas A.
We derive the line parameters (i.e., the line centroid energy,
flux and equivalent width [EW]) for these six sources and
perform a correlation analysis for the line center energies
of Cr, Mn and Fe.
Also included in the correlation analysis are
Tycho and G344.7-0.1 for which the Cr, Mn and Fe-K line parameters
were available in the literature through {\sl Suzaku} observations.
We find that the line center energies of Cr correlates very well
with that of Fe and that of Mn.
This confirms our previous findings that the Cr, Mn and Fe
are spatially co-located, share a similar ionization state,
and have a common origin in the supernova nucleo-synthesis.
We find that the ratio of the EW of the Cr emission line
to that of Fe ($\EWR\equiv \EWCr/\EWFe$)
provides useful constraints
on the SNR progenitors and on the SN explosion mechanisms:
for SNRs with $\EWR > 2\%$, a type Ia origin is favored
(e.g., N103B, G344.7-0.1, 3C\,397 and 0519-69.0);
for SNRs with $\EWR < 2\%$,
they could be of either core-collpase origin
or carbon-deflagration Ia origin.
\end{abstract}


\section{Introduction}

The ejecta in young supernova remnants (SNRs) are
usually metal-rich, and thus their X-ray spectra often
show abundant emission lines of heavy elements
(e.g., O, Ne, Mg, Si, S, Ar, Ca and Fe; see Vink 2012).
The detection of the Cr and Mn K$\alpha$ lines in the
X-ray spectra of SNRs has opened a new window to study
their progenitors and the explosion machanism of supernovae (SNe)
(Badenes et al.\ 2008a; Yang et al. 2009).
Badenes et al.\ (2008a) proposed that the metallicity of
the progenitors of Type Ia SNe can be measured from
the Mn and Cr lines in the X-ray spectra of young SNRs.
Based on this method, Badenes et al.\ (2008a,b) obtained
the metallicity of the progenitor of Tycho
and probably that of W49B as well.

Previously, we performed a Cr-K emission line survey in young SNRs
with the {\sl Chandra} archival data (Yang et al.\ 2009, hereafter
Y09). We reported the discovery of the Cr emission lines in W49B,
Cas A, Tycho and Kepler. We found a good positive correlation
between the line center energy of Cr and that of Fe which suggests a
common origin of Cr and Fe in the SN nucleosynthesis. We also
proposed that the ratio of the equivalent width (EW) of the Cr
emission line to that of Fe ($\EWR\equiv \EWCr/\EWFe$) could be used
as a valid criterion for SN classification. However, the sample
(of four SNRs) is small in size and is based on a single mission
(i.e. {\sl Chandra}). A larger sample and measurements from
other missions would be helpful for confirming our previous findings.

Due to the high sensitivity and low background above 4 keV of
the {\it X-ray Imaging Spectrometor} (XIS) onboard {\sl Suzaku}
(Koyama et al.\ 2007; Mitsuda et al. 2007), it is an ideal
instrument to study the X-ray spectra of these so-called secondary
Fe-peak elements (i.e., Cr, Mn and Fe) of SNRs. In fact, {\it
Suzaku} has detected the weak Cr and/or Mn lines in Cas A (Maeda et
al.\ 2009), Tycho (Tamagawa et al.\ 2009), Kepler (Park et al.\
2012), W49B (Ozawa et al.\ 2009), N103B (Yamaguchi \& Koyama 2010),
and marginally in G344.7-0.1 (Yamaguchi et al.\ 2012). In this work,
by using the {\it Suzaku}/XIS data, we study the Cr and Mn emission
properties of the above six SNRs and other two SNRs 3C\,397 and
0519-69.0. The new results are used to test our previous
discoveries.

In \S2 we present the spectral analyses and results.
We discuss the spatial and ionization-state correlations
between Cr and Fe and those between Cr and Mn in \S3.
In \S4 we discuss the explosion mechanism and
the progenitor properties of
Tycho, 0519-69.0, N103B, G344.7-0.1,
Kepler, 3C\,397, Cas A and W49B
based on the equivalent width ratio of Cr to Fe ($\EWR$).
A summary is given in \S5.
All through this paper, the uncertainties are given
at 90\% confidence level.

\section{Data Analysis and Results}
\subsection{Data Description and Reduction}
To study these secondary Fe-peak elements,
we have attempted to search for
an as complete as possible
set of {\sl Suzaku}/XIS data on SNRs.
To this end, we have selected six SNRs,
including Kepler, W49B, N103B and Cas A
(in which the Cr and/or Mn-K lines have
already been reported), and 3C\,397 and 0519-69.0
(in which these lines have not previously been detected).
We summarize in Table 1
the {\sl Suzaku} data employed in this paper.
The spectral analysis was not performed for Tycho
since Tamagawa et al.\ (2009) already carried out
a detailed analysis and reported the parameters of
the Cr, Mn and Fe K$\alpha$ lines.
We also note that Yamaguchi et al.\ (2012) reported
a marginal detection of the Cr and Mn lines in G344.7-0.1.
These two SNRs will be included in our analysis.
Therefore, our sample consists of eight SNRs.

\begin{table*}
\caption[]{A summary of the {\sl Suzaku} observations adopted in this work.}
\label{obsinfor}
\begin{tabular}{cccc}
\noalign{\smallskip} \hline \hline \noalign{\smallskip}
  Target      &   Obs$\_$ID     &  $t_{\rm exp}\,({\rm ks})$  &    Obs-date             \\
\noalign{\smallskip} \hline \noalign{\smallskip}
  Kepler      &   KP$^{\ast}$              &  $\sim$450       &    2010/09$\sim$2011/08 \\ \hline \noalign{\smallskip}
  W49B        &   503084010                &  $\sim$50        &    2009/03              \\
              &   504035010                &  $\sim$53        &    2009/03$\sim$04      \\ \hline \noalign{\smallskip}
  N103B       &   100013010                &  $\sim$27        &    2005/08              \\
              &   804039010                &  $\sim$210       &    2009/12$\sim$2010/01 \\ \hline \noalign{\smallskip}
  Cas A       &   100043020                &  $\sim$14        &    2006/02              \\ \hline \noalign{\smallskip}
  3C\,397       &   505008010                &  $\sim$61        &    2010/10              \\ \hline \noalign{\smallskip}
  0519-69.0   &   806026010                &  $\sim$276       &    2011/08              \\ \hline
\noalign{\smallskip} \noalign{\smallskip}
\end{tabular}

$\ast$ Key Project, Observation IDs include 505092010, 505092020, 505092030, 505092040, 505092050, 505092060, 505092070. \\
\end{table*}

The archival data are processed with HEASOFT version 6.6
and calibration database released on 2010 January 23,
following the standard criteria.\footnote{%
   http://heasarc.nasa.gov/docs/suzaku/processing/criteria\_{}xis.html.
   }

Figure 1 shows the XIS-0 images of these six SNRs.
The X-ray spectra are extracted from almost the entire remnant,
as illustrated in Figure~1.
The background spectra are extracted
from the off-source annulus regions
with an area comparable to that of the corresponding source.
For those SNRs with several observations,
we first generate the source and background spectra
as well as the corresponding response matrix (RMFs and ARFs)
for each XIS sensor and each observation,
and then combine them with {\tt addascaspec}.
Since we aim at the weak emission lines,
the spectra of the front-illuminated sensors and
that of the back-illuminated sensors\footnote{%
  We note that other than XIS-0, XIS-1 and XIS-3, XIS-2 data is only
  available for obs-ID 100013010 and 100043020.
  }
are added together in order to obtain the best statistics.
The spectra fitting is done with XSPEC version 11.3.2 (Arnaud 1996).

\begin{figure*}
\begin{minipage}[t]{0.33\textwidth}
\resizebox{5cm}{3.5cm}{\includegraphics[clip]{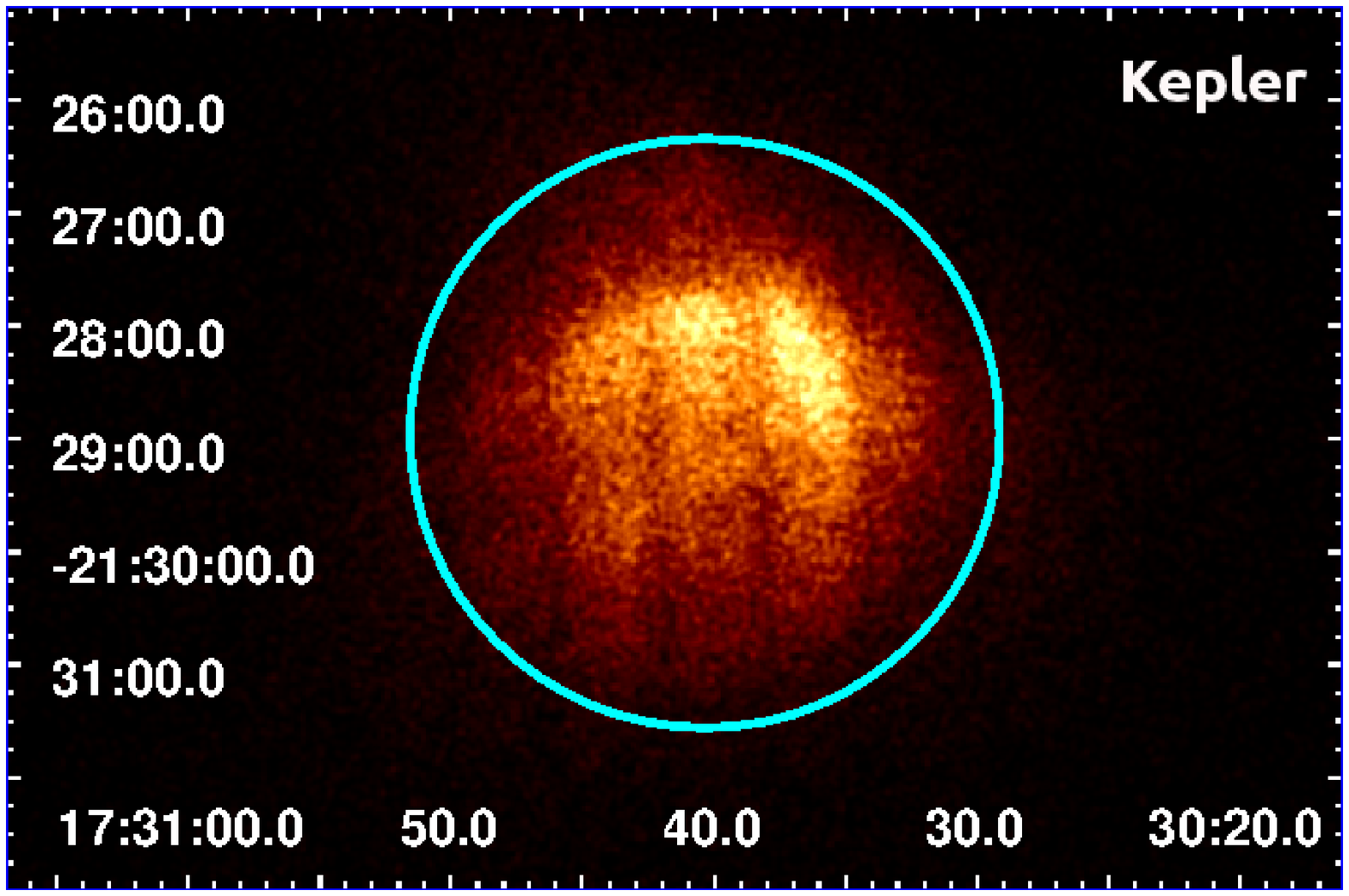}}\hspace{-0.4cm}
\resizebox{5cm}{3.5cm}{\includegraphics[clip]{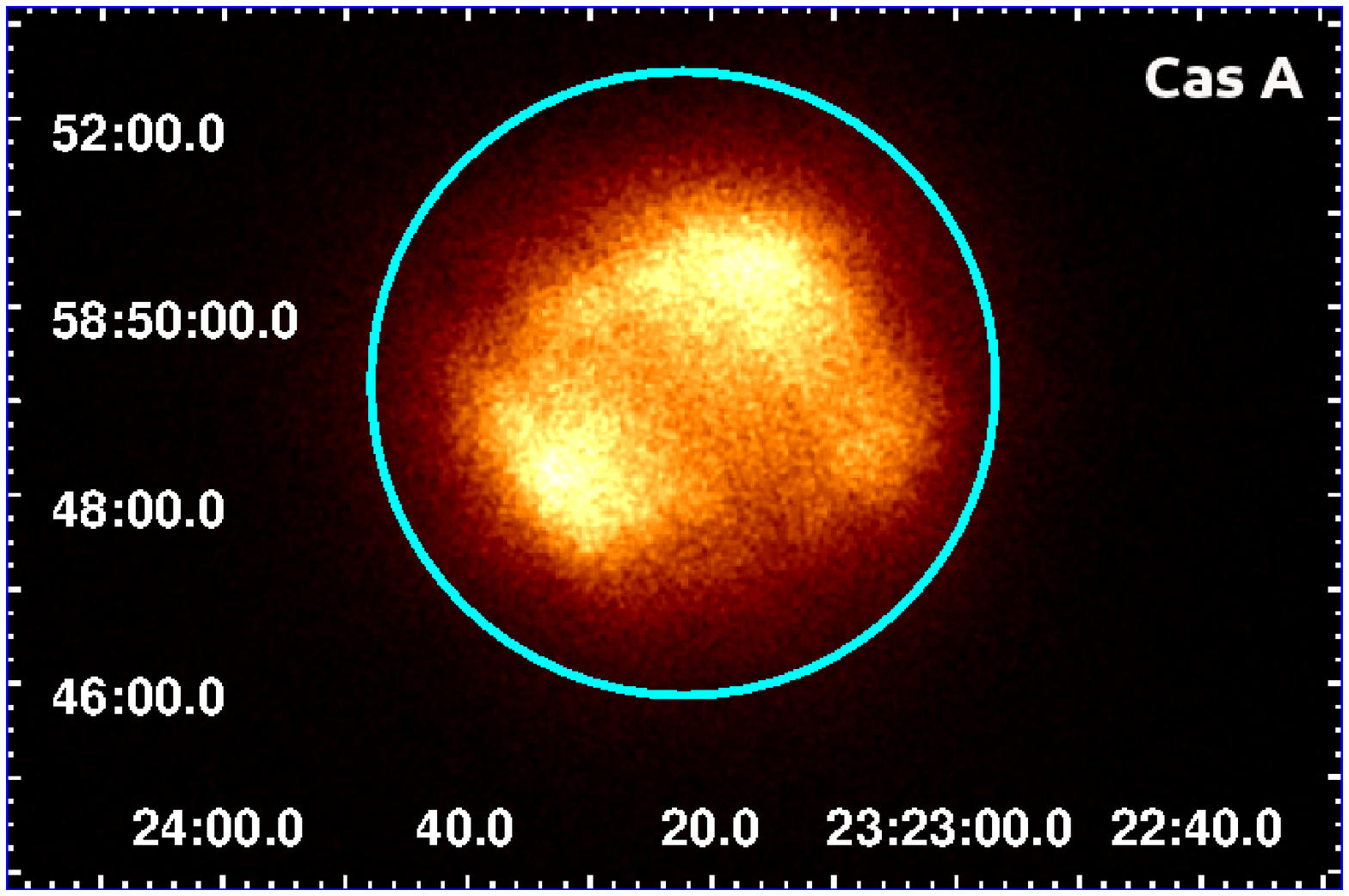}}\hspace{-0.4cm}
\end{minipage}\hspace{-0.5cm}
\begin{minipage}[t]{0.33\textwidth}
\resizebox{5cm}{3.5cm}{\includegraphics[clip]{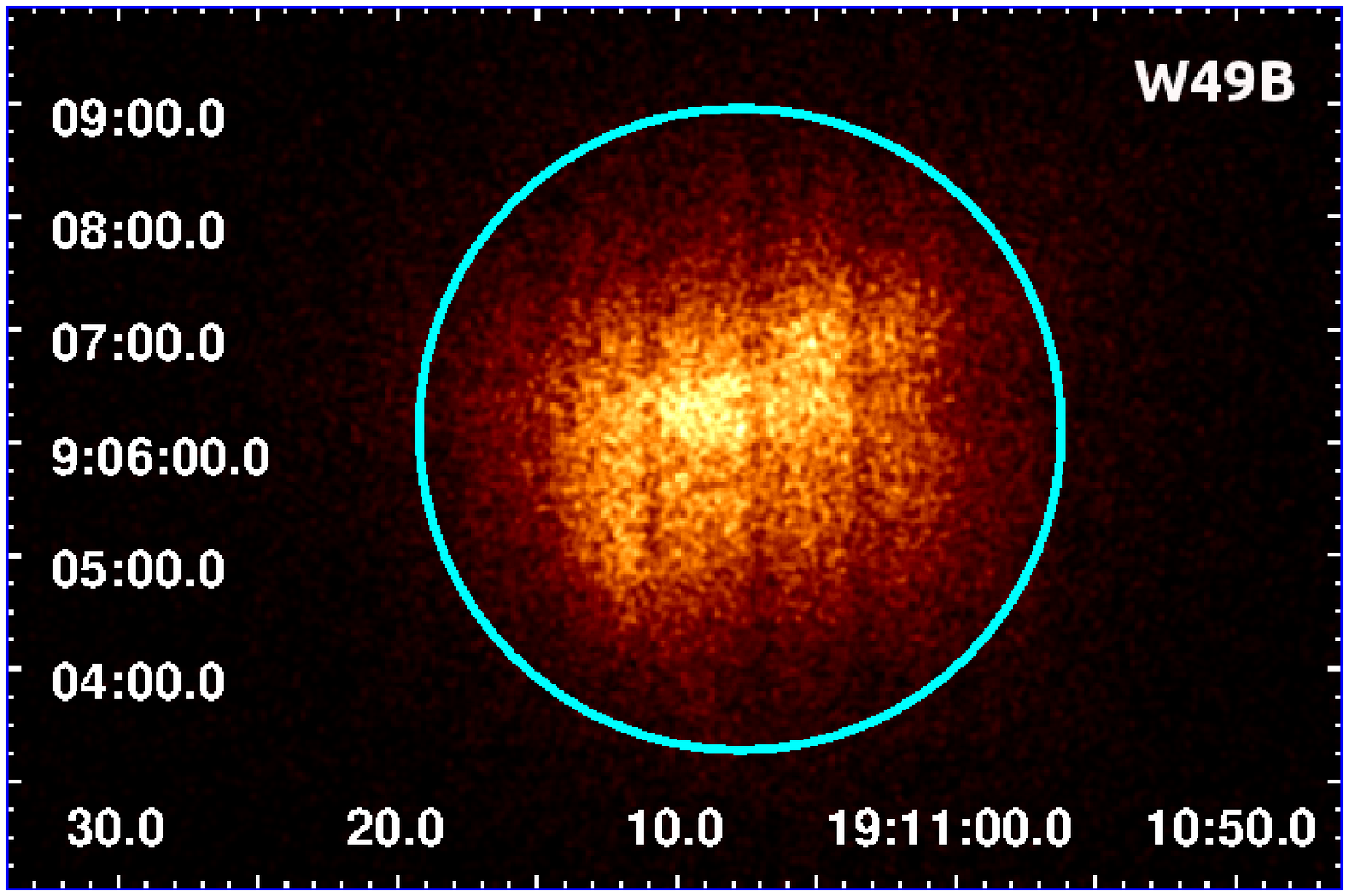}}\hspace{-0.4cm}
\resizebox{5cm}{3.5cm}{\includegraphics[clip]{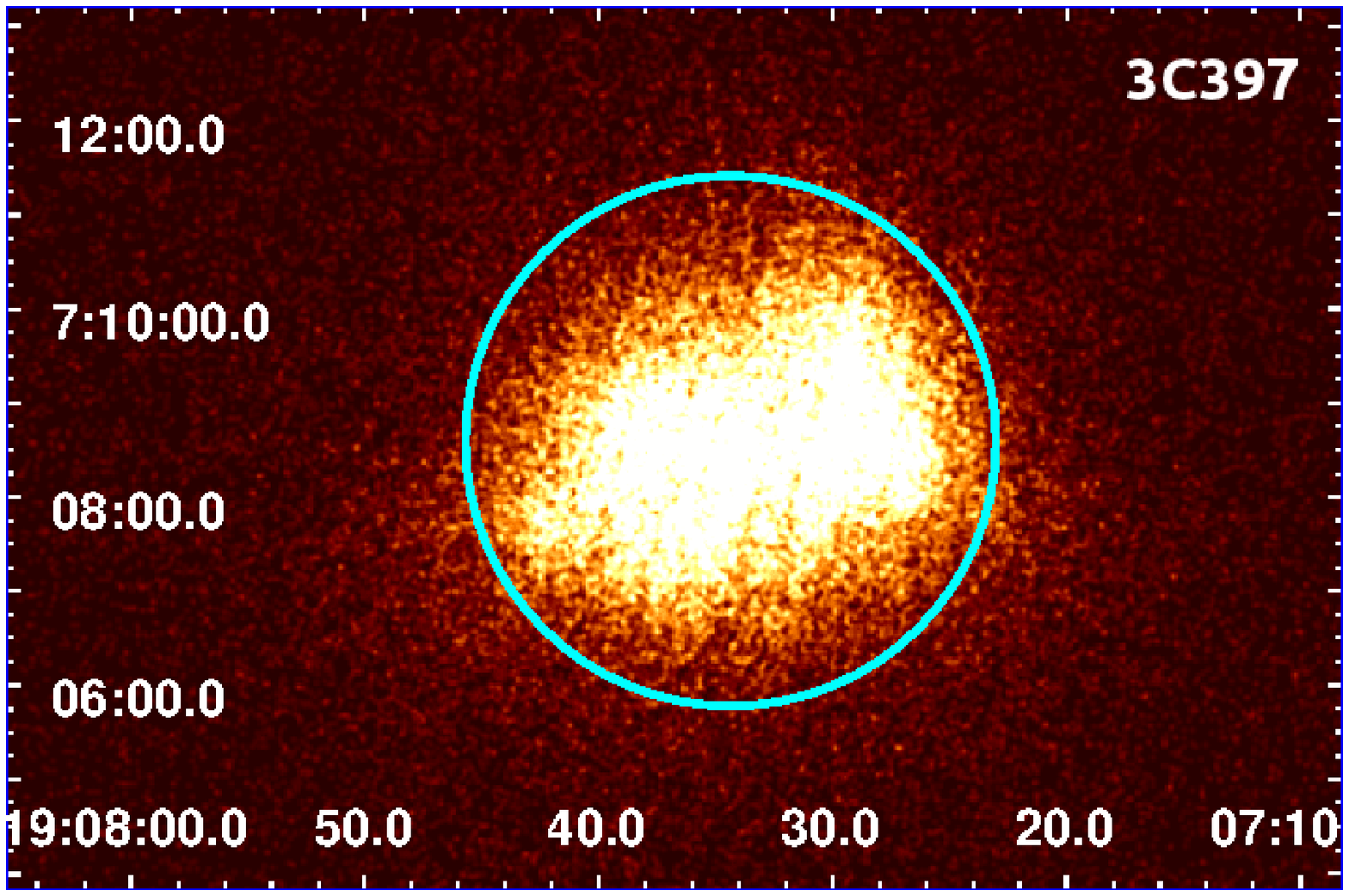}}\hspace{-0.4cm}
\end{minipage}\hspace{-0.5cm}
\begin{minipage}[t]{0.33\textwidth}
\resizebox{5cm}{3.5cm}{\includegraphics[clip]{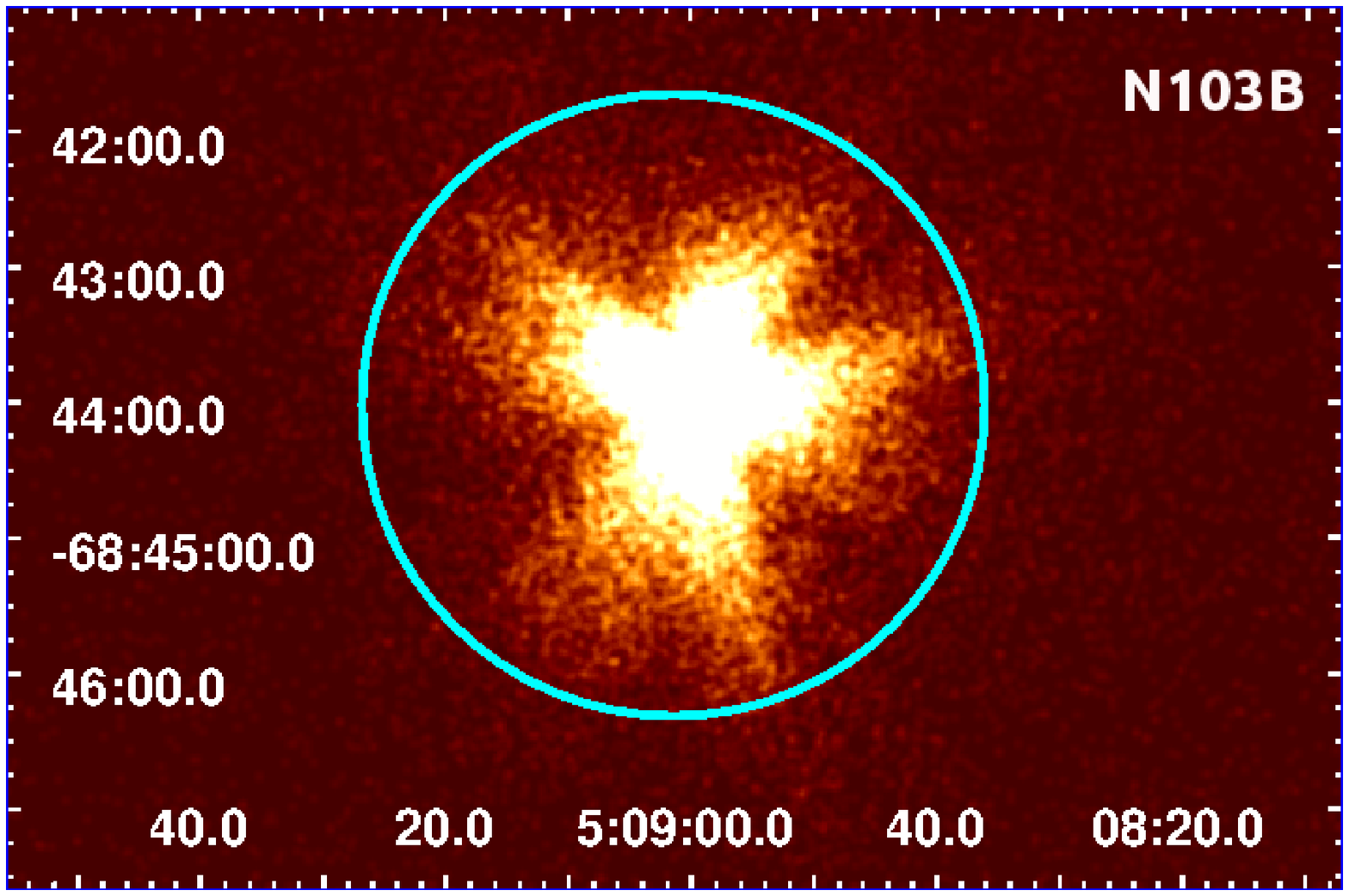}}\hspace{-0.4cm}
\resizebox{5cm}{3.5cm}{\includegraphics[clip]{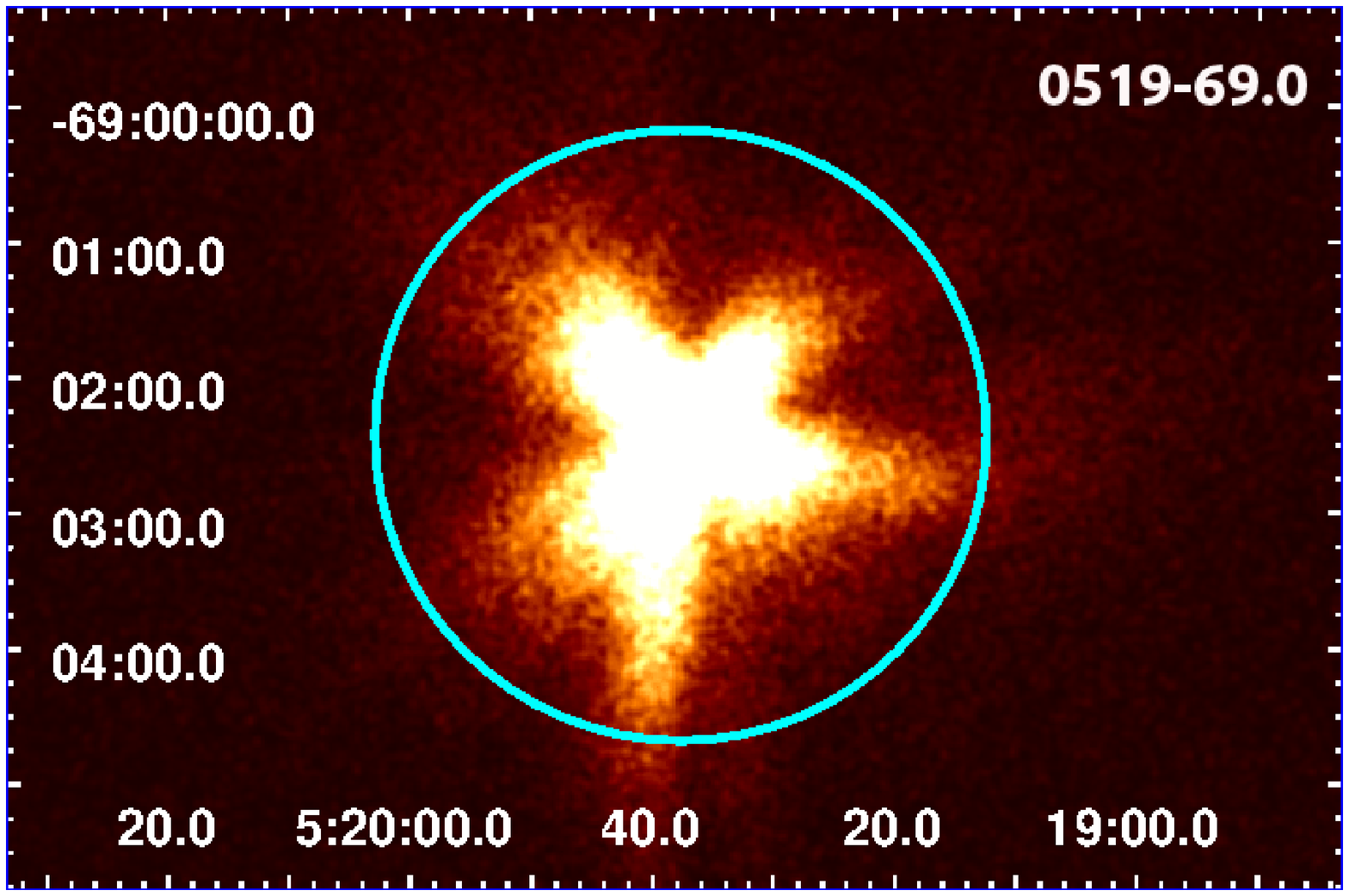}}\hspace{-0.4cm}
\end{minipage}\hspace{-0.5cm}
\caption{{\sl Suzaku} XIS0 images of Kepler, W49B, N103B, Cas A,
          3C\,397 and 0519-69.0, smoothed with gaussian function
          with radius of 3 pixels.
          The source spectral regions are overplotted in circles.}
\end{figure*}

\subsection{Spectral Analysis and Results}
The overall spectra of these six SNRs are shown in Figure~2.
Similar to Y09 in which the {\sl Chandra} data were analyzed,
we focus on the 5.0--7.0$\keV$ spectra for detailed analysis
of the Cr, Mn and Fe K-lines. All the spectra are fitted with
a power law plus multi-Gaussian components
to account for the continuum and line emissions, respectively.

\begin{figure*}
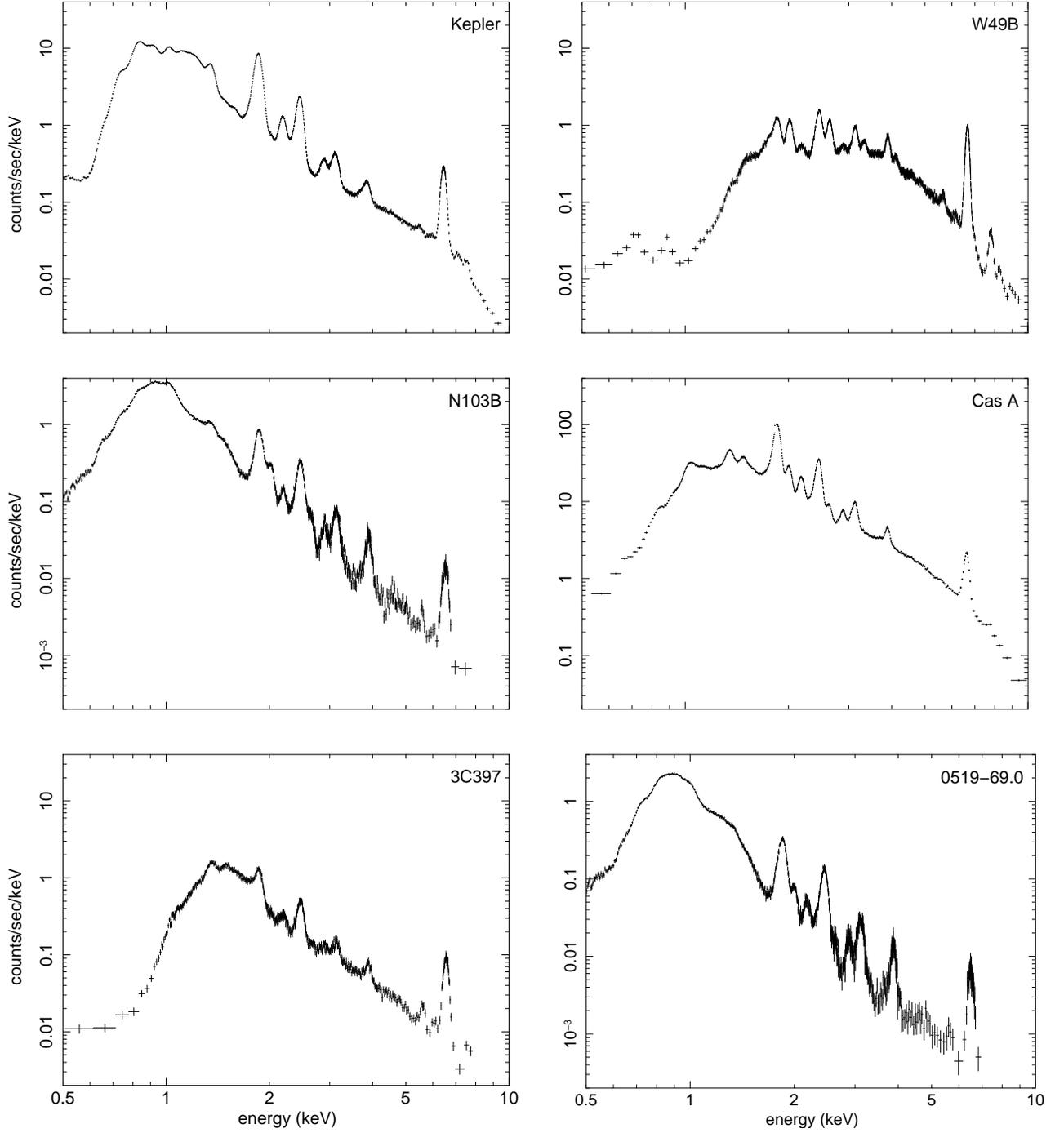

\includegraphics[width=6.1cm,angle=270,clip]{figure2_1.ps}
\includegraphics[width=6.1cm,angle=270,clip]{figure2_2.ps}
\includegraphics[width=6.1cm,angle=270,clip]{figure2_3.ps}
\includegraphics[width=6.1cm,angle=270,clip]{figure2_4.ps}
\includegraphics[width=6.1cm,angle=270,clip]{figure2_5.ps}
\includegraphics[width=6.1cm,angle=270,clip]{figure2_6.ps}
\caption{The 0.5--10.0$\keV$ spectra of Kepler, W49B, N103B,
         Cas A, 3C\,397 and 0519-69.0.}
\end{figure*}

The {\sl Suzaku} and/or {\sl Chandra} observations have already
detected the Cr and/or Mn emission lines above 5$\keV$
in Kepler (Park et al.\ 2012),
W49B (Ozawa et al.\ 2009),
N103B (Yamaguchi \& Koyama 2010)
and Cas A (Maeda et al.\ 2009).
For Kepler, W49 B and N103B,
a power law and three Gaussian components
are introduced to account for the continuum,
and the Cr, Mn and Fe line emission, respectively.
The situation for Cas A is different. We previously found that two
Gaussian components were needed to describe the {\sl Chandra} Fe-K
emission of Cas A, which might be due to the Doppler shift variation
across the remnant (Y09). This is also true for the {\sl Suzaku}
spectrum of Cas A: if we approximate the Fe-K emission line
by only one Gaussian component, the reduced $\chi^2/\dof$ is
$\sim$\,92.7/37, whereas the reduced $\chi^2/\dof$ becomes
$\sim$\,35.4/34 if we take two Gaussian components. Therefore we
take two Gaussian lines to account for the Fe-K emission of Cas A.
For these two Gaussian components, the fitted centroid energies are
$6.629\pm0.002\keV$ and $6.434\pm0.01\keV$, with their widths being
$70\pm3\eV$ and $43\pm7\eV$ respectively.

For 3C\,397 and 0519-69.0, the 5.0--7.0$\keV$ spectra
are first fitted with a power law (for the continuum emission)
plus one gaussian component (for the Fe-K line).
From the residual map,
we see one bump around 5.6$\keV$ in 0519-69.0
and two line-like structures
around 5.6$\keV$ and 6.1$\keV$ in 3C\,397.
For 3C\,397, the inclusion of a Gaussian component
near 5.6$\keV$ leads to a decrease of the reduced $\chi^2/\dof$,
from 132.6/44 to 58.1/41.
The addition of another Gaussian component near 6.1$\keV$
leads to a further decrease of the reduced $\chi^2/\dof$ to 30.4/38.
Similarly, for 0519-69.0 the reduced $\chi^2/\dof$ decreases
from 35.0/43 to 28.9/40 if a Gaussian component near 5.6$\keV$ is added.
Therefore, the 5.0--7.0$\keV$ spectrum of 3C\,397 (0519-69.0)
is fitted with a power law plus three (two) Gaussian components.
The structure near 5.6$\keV$ is due to the Cr-K line,
while the one near 6.1$\keV$ arises from the Mn-K line (see Y09).

Figure~3 shows the 5.0--7.0$\keV$ spectra along with the model
fits (as well as the residuals) of
Kepler, W49B, N103B, Cas A, 3C\,397 and 0519-69.0.
The best fit parameters of the Cr, Mn and Fe-K emission lines
are listed in Table~2.
For Cas A in which the Fe-K line is fitted in terms of
two Gaussian components, the centroid energy is
the emission-weighted mean value, while the flux and EW
are the sums of the two Gaussian lines.
For Tycho and G\,344.7-0.1, the parameters are taken from
Tamagawa et al.\ (2009) and Yamaguchi et al.\ (2012).

\begin{sidewaystable}[h]
\caption{The Cr, Mn and Fe line parameters of Tycho, Kepler, W49B, N103B, Cas A, G344.7-0.1, 3C\,397 and 0519-69.0.}
\label{CrMnFe}

\begin{tabular}{cccccccccc}
\noalign{\smallskip} \hline \hline \noalign{\smallskip}
\multicolumn{2}{c}{             }   &       Kepler           &   W49B                   &   N103B                  &            Cas A                  &         3C\,397               &     0519-69.0         &   Tycho$^\dag$    &   G344.7-0.1$^\ddag$ \\
 \noalign{\smallskip} \hline \noalign{\smallskip}
\multicolumn{2}{c}{Type}                          &          Ia\,(?)           &        C-C$^d$\,(?)      &           Ia\,(?)                  &            C-C       &          Ia\,(?)         &          Ia                &            Ia            &     Ia\,(?)
\\ \noalign{\smallskip} \hline \noalign{\smallskip}
\multirow{3}{*}{Cr}    &   centroid$^a$           &       5490$\pm{20}$      &  5655$^{+6}_{-10}$        &  5586$^{+25}_{-32}$     &  5610$^{+28}_{-56}$     &  5613$^{+24}_{-39}$      &    5644$^{+80}_{-78}$    &     5480$\pm{20}$         &  5526$^{+70}_{-66}$            \\ \noalign{\smallskip}
                       &         flux$^b$         &  0.50$^{+0.08}_{-0.10}$  &  4.5$^{+0.34}_{-0.35}$    & 0.16$^{+0.09}_{-0.06}$  &   3.4$^{+1.3}_{-0.5}$        &  1.46$^{+0.31}_{-0.29}$  &  0.08$^{+0.07}_{-0.04}$   &   2.45$^{+0.48}_{-0.42}$        &   0.36$^{+0.16}_{-0.14}$
\\ \noalign{\smallskip}
                       &     EW$^c$               &  22.7$^{+4.5}_{-4.8}$    &  96.3$^{+6.3}_{-6.5}$     &  116$^{+58}_{-40}$      &   8.6$^{+2.7}_{-1.2}$        &  198$^{+40}_{-39}$       & 206$^{+200}_{-100}$       &  23.8$^{+8.5}_{-13.0}$          &       $153$
\\  \noalign{\smallskip} \hline \noalign{\smallskip}
\multirow{3}{*}{Mn}    &       centroid$^a$       &  5960$^{+40}_{-30}$      &  6161$\pm{20}$            &  6020$^{+70}_{-75}$     &      $-$                    &   6084$^{+53}_{-64}$     &      $-$                 &        5950$\pm{50}$      &   6085$^{+107}_{-120}$                       \\ \noalign{\smallskip}
                       &       flux$^b$           &  0.27$^{+0.18}_{-0.02}$  &  1.5$^{+0.1}_{-0.6}$      &  0.10$\pm{0.06}$        &
        $-$            &  0.88$^{+0.37}_{-0.29}$  &      $-$                 &  0.24$^{+0.12}_{-0.13}$   &        1.1$\pm{0.4}$
\\ \noalign{\smallskip}
                       &         EW$^c$           &   15.0$^{+10.0}_{-2.9}$  &   44.7$^{+3.0}_{-18.0}$   & 84.5$\pm{50.0}$         &
        $-$            &   158$^{+75}_{-80}$      &           $-$            &     13.7$^{+13.5}_{-11}$  &           $106$
\\ \noalign{\smallskip} \hline \noalign{\smallskip}
\multirow{3}{*}{Fe}    &   centroid$^a$           &      6448$\pm{1}$        &      6662$\pm{2}$         &  6528$\pm{7}$           &   6613$^{+2}_{-1}$       &  6556$^{+5}_{-4}$        &  6496$^{+9}_{-12}$       &       6445$\pm{1}$        &   6447$^{+11}_{-12}$           \\ \noalign{\smallskip}
                       &      flux$^b$            &   39.9$^{+0.2}_{-0.4}$   &       127$^{+1}_{-2}$     &   3.3$\pm{0.2}$         &
   247$^{+4}_{-3}$     &   18.8$^{+0.7}_{-0.6}$   &        1.3$\pm{0.1}$     &     69.1$^{+0.6}_{-0.9}$  &        2.8$\pm{0.3}$
\\ \noalign{\smallskip}
                       &         EW$^c$           &   2680$^{+14}_{-26}$     &  5378$^{+42}_{-80}$       &  4370$^{+220}_{-190}$   &
   871$^{+12}_{-10}$   &   4490$^{+167}_{-143}$   &      4930$\pm{380}$      &     1040$^{+30}_{-20}$    &           $1758$
\\  \noalign{\smallskip} \hline  \noalign{\smallskip}
\multirow{2}{*}{$\EWR$}&    {\sl Suzaku}          & 0.85$^{+0.16}_{-0.17}$\% &     1.8$\pm{0.1}$\%       &  2.7$^{+1.2}_{-0.8}$\%  & 0.99$^{+0.30}_{-0.15}$\% &     4.4$\pm{0.7}$\%    &   4.2$^{+3.7}_{-1.7}$\%  &    2.3$^{+0.8}_{-1.2}$\%  &         $\sim8$\%
\\ \noalign{\smallskip}
                       & {\sl Chandra}$^e$        & 0.63$^{+0.30}_{-0.25}$\% & 1.6$^{+0.40}_{-0.36}$\%   &           $-$          &
0.85$^{+0.18}_{-0.07}$\% &          $-$           &            $-$           &  3.6$^{+3.2}_{-1.1}$\%    &       $-$
\\ \noalign{\smallskip} \hline \noalign{\smallskip}

\end{tabular}

$^a$  Line center in unit eV          \\
$^b$  Flux in unit: $10^{-5}\photons\cm^{-2}\s^{-1}$     \\
$^c$  Eqivalent Width in unit eV      \\
$^d$  C-C stands for core-collapse            \\
$^\dag$ Tamagawa et al.\ 2009     \\
$^\ddag$ Yamaguchi et al.\ 2012.
         The EWs are from private communications
         with H. Yamaguchi \\
$^e$ The {\sl Chandra} $\EWR$ values are taken from Y09

\end{sidewaystable}

\begin{figure*}
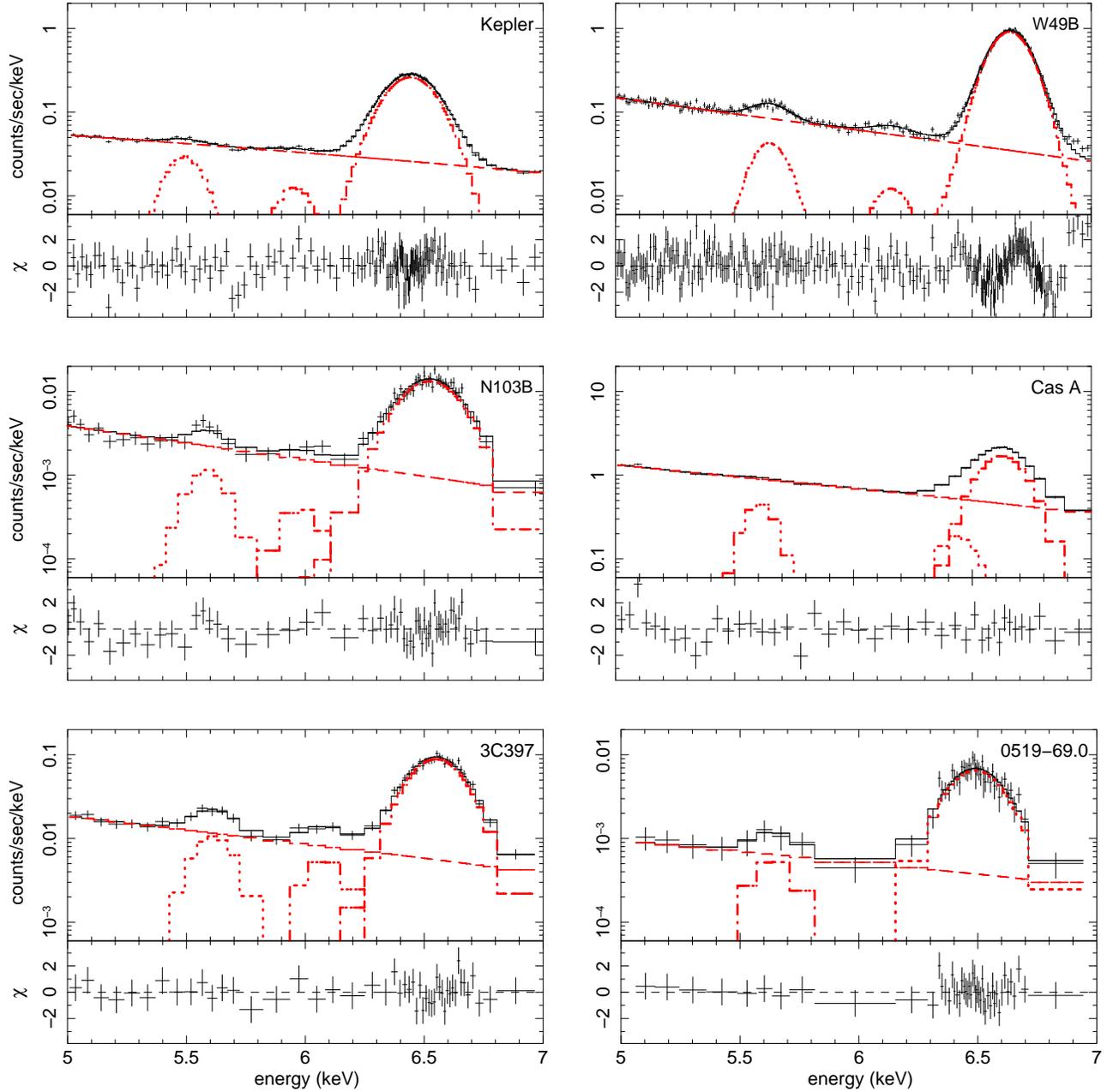

\includegraphics[width=5.6cm,angle=270,clip]{figure3_1.ps}
\includegraphics[width=5.6cm,angle=270,clip]{figure3_2.ps}
\includegraphics[width=5.6cm,angle=270,clip]{figure3_3.ps}
\includegraphics[width=5.6cm,angle=270,clip]{figure3_4.ps}
\includegraphics[width=5.6cm,angle=270,clip]{figure3_5.ps}
\includegraphics[width=5.6cm,angle=270,clip]{figure3_6.ps}
\caption{The 5.0--7.0$\keV$ spectra of Kepler, W49B, N103B,
         Cas A, 3C\,397 and 0519-69.0. The lines plot
         the model fits (see \S3). The Cr line for Cas A is
         enhanced by a factor of 10,
         while that for Kepler a factor of 5. }
\end{figure*}

\section{Spatial and Ionization State Correlations
         between Cr, Mn and Fe}
The Cr, Mn and Fe-K line parameters
derived here from {\sl Suzaku}
for Kepler, W49B, N103B and Cas A (Table~3)
are generally consistent with those
from {\sl Chandra} (see Table~2 in Y09)
and/or previous studies from {\sl Suzaku}
(Tamagawa et al.\ 2009;
Ozawa et al.\ 2009;
Yamaguchi \& Koyama 2010;
Park et al.\ 2012;).
We suggest that the different line center energies
of the Cr-K lines in these SNRs reflect the different
ionization states of Cr (Y09).
The Cr line center energies in N\,103B and G\,344.7-0.1
reveal their ionization states to be Be- and B-like,
respectively (Hata \& Grant 1984; also see Table~4 in Y09).
The Cr line centroids of 3C\,397 and 0519-69.0 are higher than
those of N\,103B or G\,344.7-0.1, implying higher ionization
states (i.e., Li- and He-like).

In Figure ~4, we plot the line center energies of Fe-K
against those of Cr-K in all eight SNRs listed in Table~2.
It is seen that these two sets of line center energies
closely correlate, with a correlation coefficient of $r\approx0.79$.
This further confirms that Cr and Fe are basically
in similar ionization states in these SNRs.
We note that such a close correlation was already seen in Y09,
but for a smaller sample (of 4 SNRs).

\begin{figure*}
\includegraphics[width=10cm,clip]{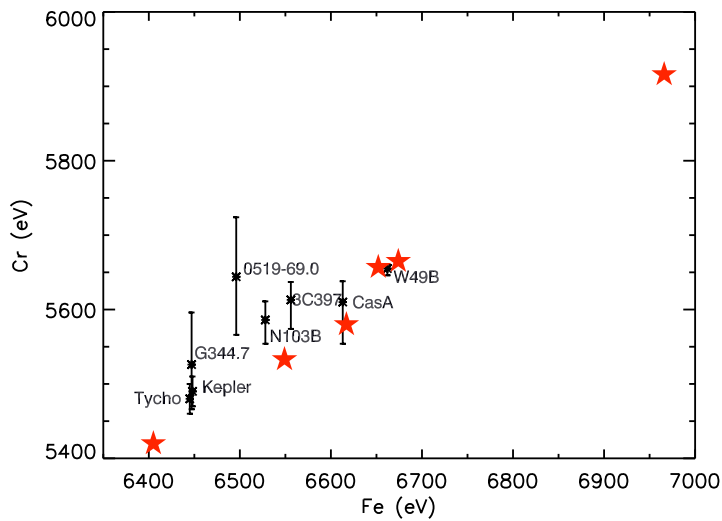}
\caption{The line center energies of Cr versus Fe
         in Tycho, Kepler, W49B, N103B, Cas A, G344.7-0.1,
         3C\,397 and 0519-69.0. The errors of the Fe line
         are not shown as they are almost within the size
         of the symbols.
         The stars represent the line center energies
         of Cr and Fe in various ionization states
         (i.e., from H-like to B-like as well as neutral states,
          c.f. Table 4 in Y09).
         }
\end{figure*}

Thanks to the high sensitivity of {\sl Suzaku}/XIS, the Cr and Mn-K
lines are for the first time simultaneously detected in several SNRs
of this sample, including Tycho, Kepler, W49B, N\,103B, G\,344.7-0.1
and 3C\,397. In Figure~5 we plot the line center energy of Cr
against that of Mn for these SNRs. It is apparent that there exists
a positive correlation between the centroid energies of these two
lines, with a correlation coefficient of $r\approx0.85$. Since the
line center energy is closely related to the ionization age of the
emitting plasma, such a positive correlation implies that Cr and Mn
also share a similar ionization age, just as Cr and Fe in the SNRs.

Based on the nucleosynthesis models
(Nomoto et al.\ 1984; Woosley \& Weaver 1994; Woosley et al.\ 1995),
the most abundant isotope of chromium (i.e. $^{52}$Cr)
is the product of $^{52,53}$Fe decay
during explosive silicon burning,
whereas the single isotope of manganese, $^{55}$Mn,
is produced mostly in explosive silicon burning
and nuclear statistical equilibrium as $^{55}$Co.
The most abundant Fe, $^{56}$Fe, is decayed from $^{56}$Ni,
which is also produced mainly from explosive silicon burning.
Since Cr, Mn and Fe are all made
in the same nucleosynthesis process,
it is natural that they located near each other
and thus in a similar ionization state.

\begin{figure*}
\includegraphics[width=10cm,clip]{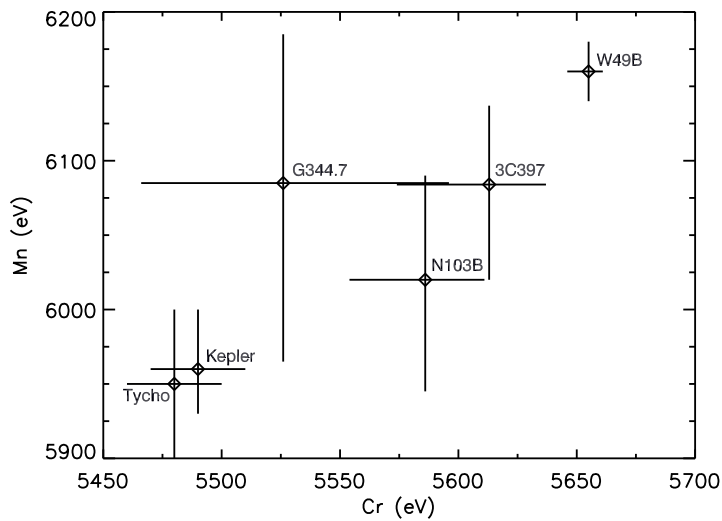}
\caption{The line center energies of Cr versus Mn
         in Tycho, Kepler, W49B, N103B, 3C\,397 and G344.7-0.1.
         }
\end{figure*}

\section{Can SN Explosion Be Classified According to $\EWR$?}

Theoretically, the production of Cr and Fe varies with the
nature of the SN progenitors and the explosion mechanisms.
Generally speaking, core-collapse SNe cannot produce as much Cr as
type Ia SNe, and the theoretical Cr-to-Fe mass ratio
$\MCr/\MFe$ is often less than 2\% for core-collapse SNe
(Woosley et al.\ 1995, Thielemann et al.\ 1996, Maeda et al.\ 2003).
Theoretical calculations also suggest that the more massive
a progenitor is, the more Cr it will produce (Nakamura et al. 1999).
%
We note that these calculations have not taken neutrino process
into account. This could affect the yields of the Fe group nuclei.
With the neutrino process taken into account,
Fr\"{o}hlich et al.\ (2006) found that the abundances of heavy
elements, especially Co, Ni, Cu, Zn, with respect to Fe, are
appreciably different from those calculated by Thielemann et al.\ (1996).
However, the relative abundances of Cr and Mn to Fe are generally
in agreement (see Figure~8 of Fr\"{o}hlich et al.\ 2006).
%

The Cr-to-Fe mass ratio $\MCr/\MFe$ of Type Ia SNe differs
significantly among SNe of different explosion mechanisms: small
$\MCr/\MFe <1\%$ for carbon deflagration explosion and large
$\MCr/\MFe >1.5\%$ for delayed-detonation explosion (Nomoto et al.\
1997, Iwamoto et al.\ 1999). For the delayed-detonation model,
$\MCr/\MFe$ decreases as the transition density increases.
%
We note that the synthesized masses of Cr and Fe predicted from
the two-dimensional simulations on delayed-detonation models
as well as the two- and three-dimensional simulations
on pure deflagration models (Travaglio et al.\ 2004, Maeda et al.\
2010) could differ by a factor of $\simali$2
from those predicted from the one-dimensional simulations
(Nomoto et al.\ 1997, Iwamoto et al.\ 1999).
However the mass ratio ($\MCr/\MFe$) is generally consistent
with each other, i.e, M(Cr)/M(Fe) $<$1\% for
pure deflagration models and $>$1.5\% for
delayed-detonation models.
%

Based on {\sl Chandra} data,
we propose that one could use $\EWR$ of SNRs to constrain the
the corresponding SN explosion mechanisms.
As discussed in Y09 and earlier in this paper, Cr and Fe are spatially
co-located, and thus share the same ionization state, temperature
and ambient electron density.
Therefore, $\EWR$, the ratio of the EW
of the Cr emission line to that of Fe should reflect the
corresponding mass ratio ($\MCr/\MFe$).
%

From Table 2 we can see that the available $\EWR$ ratios
from {\sl Chandra} are generally consistent with
the corresponding {\sl Suzaku} values
within the confidence range.
In the following we will further explore this,
based on {\sl Suzaku} data.

As a single-degenerate Type Ia SNR (Lu et al.\ 2011),
Tycho is considered to originate from a delayed detonation explosion
with a relatively small transition density,
based on the EW ratio of Cr to Fe
$\EWR\approx 3.6^{+3.2}_{-1.1}\%$ (Y09).
The {\sl Suzaku} value of $\EWR\approx 2.3^{+0.8}_{-1.2}\%$
derived here is generally consistent with that of {\sl Chandra}
within uncertainty, supporting the delayed-detonation explosion
scenario for the explosion mechanism of Tycho's progenitor,
with the transition density $\sim 2.2\times10^7\g\cm^{-3}$
(Nomoto et al.\ 1997; Iwamoto et al.\ 1999).

SNR 0519-69.0 is confidently known to originate from SN Ia
based on its light echoes (Rest et al.\ 2005, 2008)
and X-ray spectra (Hughes et al.\ 1995).
Recent studies show that it may originate
either from a supersoft source of single-degenerate scenario
or from a double-degenerate system (Edwards et al.\ 2012),
with the latter (to our knowledge) not well studied
in current theoretical calculations.
Its large Cr-to-Fe EW ratio $\EWR\approx4.2\%$
favors the delayed-detonation scenario
with transition density of $\sim1.5\times10^7\g\cm^{-3}$.

N103B tends to be classified as
a remnant of SN Ia (Badenes et al.\ 2007 and reference therein),
although a core-collapse origin cannot be ruled out
(van der Heyden et al. 2002).
However, {\sl ASCA} observations show that it has strong
K$\alpha$ lines from Si, S, Ar, Ca,
and the calculated nucleosynthesis yields are
qualitatively inconsistent with
the core-collapse scenario (Hughes et al.\ 1995).
Based on {\sl Chandra} ACIS data,
Lewis et al.\ (2003) found that the distribution
of the ejecta and the yields of the intermediate-mass species
are consistent with the model predictions for Type Ia events.
Such a (Type Ia) classification is further
supported by the emission morphology of
the X-ray lines (Lopez et al.\ 2009).\footnote{%
   Nevertheless, the {\sl XMM-Newton} and {\sl Chandra} spectra
   show a high elemental abundance of O and Ne and low abundance
   of Fe, which implies that N103B might originate from
   a type II SN rather than type Ia (van der Heyden et al.\ 2002).
   }
The EW ratio of Cr-to-Fe of N103B, $\EWR\approx 2.7\%$,
is similar to (but somewhat slightly larger than) that of Tycho.
Since theoretical models predict the mass ratio of Cr-to-Fe
often to be $\MCr/\MFe <2\%$ for core-collapse  SNe
(Woosley et al.\ 1995, Thielemann et al.\ 1996, Maeda et al.\ 2003),
we favor the SN Ia origin for N103B.
If this is indeed the case,
such a $\EWR$ ratio would imply that
a delayed-detonation model would be required
during its explosion,
and the transition denstity would be similar to
(but slightly smaller than) that of Tycho,
since the theoretical calculations suggests that
the $\MCr/\MFe$ ratio decreases as the transition
density increases (Nomoto et al.\ 1997,
Iwamoto et al.\ 1999, Travaglio et al.\ 2004, 2005).

G344.7-0.1 is suggested to be a core-collapse SNR from many
observational results. There appears to be a point-like source at its
geometrical center, although it might be a foreground object (Combi
et al. 2010). The SNR is also associated with a nearby molecular
cloud or a wind-blown bubble (Combi et al. 2010; Giacani et al.
2011). Furthermore, its highly asymmetric X-ray line emission
morphology is similar to other core-collapse SNRs (Lopez et al.
2011). However, Yamaguchi et al. (2012) presented an X-ray
spectroscopic study of this SNR using {\sl Suzaku}, which favors a
type Ia origin. They found that its abundance pattern is highly
consistent with that expected for a somewhat-evolved type Ia SNR. They
further indicated that G344.7-0.1 is the first possible type Ia
SNR categorized as a member of the so-called ``mixed-morphology'' class.
Its large $\EWR \sim 8\%$ also supports an Ia origin (Woosley et al. 1994).
Nevertheless, we note here that the detection of the Cr-K line in
G344.7-0.1 is marginal and the line paramters are with large
uncertainties (Yamaguchi et al.\ 2012). Deeper observation of this
SNR would lead to better-constrained line parameters and thus
a clarification of its origin.

For Kepler, the O/Fe ratio observed
in the X-ray spectrum (Reynolds et al.\ 2007)
and the X-ray line emission morphology (Lopez et al.\ 2011)
favors a Type Ia origin.
More recently, Patnaude et al.\ (2012) performed hydrodynamical
and spectral modeling to constrain the origin of the Kepler SNR.
They found that the delayed-detonation model interacting with
a wind provides a good match both spectrally and dynamically.
If we assume a Type Ia origin for Kepler, its small EW ratio
of Cr-to-Fe ($\EWR\approx0.8\%$) would imply a carbon deflagration
model without detonation.

The classification of 3C\,397 is not conclusive.
Chen et al.\ (1999) argued that the cloudlet environment
around this SNR favors a type Ia origin.
They also suggested that the bipolar bubble structure
seen in its X-ray image might be formed through mass accumulation,
which would be the case for a SN Ia progenitor of
mass-losing binary system. Meanwhile, a compact object
is not detected, either in X-ray or in radio (Safi-Harb et al.\ 2005).
The large EW ratio of Cr-to-Fe of $\EWR\approx4.4\%$
derived here also favors a type Ia classification,
as core-collapse SNe often yield relatively small mass ratio
of Cr-to-Fe ($\MCr/\MFe <2\%$).
As discussed above, the large $\EWR$ ratio of 3C\,397
requires detonation.
With its $\EWR$ ratio being similar to SNR\,0519-69.0,
3C\,397 may also have a transition density similar to SNR\,0519-69.0
($\sim$\,$1.5\times10^7\g\cm^{-3}$).
Based on multiwavelength imaging and spectral studies,
Safi-Harb et al.\ (2005) argued that 3C\,397 will evolve
into a mix-morphology SNR. If so, 3C\,397 would be another
type Ia in the ``mixed-morphology'' SNR category,
just like G344.7-0.1.

For the core-collapse SNR Cas A,
its $\EWR\approx 0.99\%$ ratio supports a progenitor mass
of $\sim$\,15--25\,$M_{\odot}$ and an asymmetric explosion scenario.

W49B tends to be classified as a core-collapse SNR
(Miceli et al.\ 2006; Keohane et al.\ 2007).
This is supported by Y09 from its EW ratio $\EWR\approx1.6\%$
and by Lopez et al.\ (2009) from its X-ray line emission morphology.
Ozawa et al.\ (2009) detected the over-ionized plasma in W49B
with {\sl Suzaku}. They argued that a massive progenitor that
had blown a stellar wind would be favored,
if the origin of the plasma is via the cooling caused
when the blast wave breaks out of some ambient matter into
the rarefied interstellar medium.
Nevertheless, Hwang et al.\ (2000) suggested a type Ia origin
based on the relative abundance of Mg, S, Ar, Ca, Fe and Ni to S.
Badenes et al.\ (2008b) suggested that if one assumes a Type Ia
explosion mechanism for W49B, its Mn/Cr ratio could imply
a metalicity of $0.041^{+0.056}_{-0.036}$ for its progenitor.

Its $\EWR$ ratio of $\sim$\,1.8\% derived here suggests
either a core-collapse origin or a SN Ia of delayed-detonation explosion.
Woosley et al.\ (1995) found that the more massive
a progenitor is, the more amount of Cr will be produced.
Assuming W49B to be of core-collapse origin,
we would expect a much larger progenitor mass for W49B
compared to that of Cas A, as the EW of the Cr emission
line (often considered as a valid representation of the Cr abundance)
of W49B is ten times that of Cas A.
If one assumes a SN Ia origin, its $\EWR$ ratio
(of $\sim$\,1.8\%) implies a relatively large
deflagration-detonation transition density,
$\sim3\times10^7\g\cm^{-3}$.

Finally, we note that the above discussions are based on the overall mass ratio of Cr
to Fe in SNRs. The X-ray emission from young SNRs is however
prominent only above the reverse shock,
[e.g., in Tycho, Warren et al.\ (2005) assumed that
the Fe-K$\alpha$ emission originates from the inner
most position of the shocked ejecta and located
the reverse shock at its inner edge;
in Cas A, Gotthelf et al.\ (2001)
and Helder \& Vink (2008)
placed the reverse shock at the location
where a sharp rise in X-ray emissivity
with increasing radius occurs].
As discussed in \S3 and \S4.1 of Y09, Cr and Fe are
in similar ionization states and well mixed, so the observed $\EWR$
ratio should be a good indicator of the overall mass ratio of Cr to Fe
for the remnant as a whole.
Therefore, our conclusion should be reliable
no matter where the reverse shock is and
how much ejecta have been overtaken.
%

\section{Summary}
We perform a {\sl Suzaku} X-ray spectroscopic analysis
of the Cr, Mn and Fe-K emission lines in young SNRs.
The principal results of this work are the following:
\begin{enumerate}
\item The detection of the Cr and/or Mn lines are
      reported, for the first time, in 3C\,397 and 0519-69.0,
      and confirmed in Kepler, W49B, N103B and Cas A.

\item The line parameters (i.e., the centroid energy, flux and
      equivalent width) are derived for these six sources.

\item We perform a correlation analysis between the line center
      energies of Cr, Mn and Fe-K for these six sources together
      with Tycho and G344.7-0.1 of which the Cr, Mn and Fe-K line
      parameters are available in the literature.
      We find a positive correlation between the Cr and Fe-K line
      center energies, as found previously in Y09 based on
      {\sl Chandra} data of four SNRs.
      Such a positive correlation is also found between the Cr and
      Mn-K line centroids.
      This supports the common origin of Cr, Mn and Fe
      in the SN nucleosynthesis, and suggests that they
      are spatially co-located.

\item The EW ratio of Cr-to-Fe ($\EWR$),
      a good representation of the Cr-to-Fe mass ratio
      ($\MCr/\MFe$) of the SNR,
      provides useful constraints on its progenitor
      and the SN explosion mechanism.
      For Tycho (Type Ia), 0519-69.0 (Type Ia)
      and Cas A (core-collapse),
      the derived EW ratios $\EWR$ are consistent with
      their classification.
      The large $\EWR$ ratios derived for N103B, G344.7-0.1 and 3C397
      suggest a Type Ia SN explosion as their origins.
      For the putative type Ia SNRs Kepler,
      its small $\EWR$ ratio suggests either a carbon
      deflagration explosion or a core-collapse origin
      (which needs to be further explored).
      For W49B, its $\EWR$ ratio suggests that either it is of
      core-collpase origin with a very massive progenitor,
      or it is of Type Ia origin with a relatively large
      detonation-deflagration transition density.
\end{enumerate}

\begin{acknowledgements}
We thank the anonymous referee for helpful comments.
AL and XJY are supported in part by a NASA/Chandra theory grant
and NSF/AST 1109039.
This research made use of data obtained from the {\sl Suzaku}
satellite, a collaborative mission between the space agencies of
Japan (JAXA) and the USA (NASA). We thank Satoru Katsuda for his help
on {\sl Suzaku} data analysis and H. Yamaguchi for providing the
EWs of the Cr, Mn and Fe-K lines of G344.7-0.1.
This work is also supported by the Natural Science Foundation of China
through grants 10903007, 11273022, 11133002, 11233001,
and by JSPS KAKENHI Grand Number 23000004.
\end{acknowledgements}

\end{document}